\documentclass[fleqn,usenatbib]{mnras}
\usepackage{newtxtext,newtxmath}
\usepackage[T1]{fontenc}
\usepackage{ae,aecompl}
\usepackage{graphicx}
\usepackage{amsmath}
\newcommand{\angstrom}{\text{\normalfont\AA}}
\usepackage{lipsum}
\usepackage{multirow}
\usepackage[section]{placeins}
\usepackage{mathrsfs}
\usepackage{ulem}
\usepackage[dvipsnames]{xcolor}
\usepackage[caption=false]{subfig}

\newcommand\lsim{\mathrel{\rlap{\lower4pt\hbox{\hskip1pt$\sim$}}
        \raise1pt\hbox{$<$}}}
\newcommand\gsim{\mathrel{\rlap{\lower4pt\hbox{\hskip1pt$\sim$}}
        \raise1pt\hbox{$>$}}}

\title[EM counterparts of {\it LISA} binaries]{Identifying 
 the electromagnetic counterparts of {\it LISA} massive black hole binaries in archival LSST data}

\author[Xin \& Haiman]{Chengcheng~Xin,$^{1}$
Zolt{\'{a}}n~Haiman,$^{1,2}$ \\
$^{1}$Department of Astronomy, Columbia University, New York, NY, 10027, USA\\
$^{2}$Department of Physics, Columbia University, New York, NY, 10027, USA\\}

\date{Accepted XXX. Received YYY; in original form ZZZ}

\pubyear{2024}

\begin{document}
\label{firstpage}
\pagerange{\pageref{firstpage}--\pageref{lastpage}}
\maketitle

\begin{abstract}
The Vera C. Rubin Observatory’s Legacy Survey of Space and Time (LSST) will catalogue the light-curves of up to 100 million quasars. Among these there can be up to approximately 100 ultra-compact massive black hole (MBH) binaries, which 5-15 years later can be detected in gravitational waves (GWs) by the Laser Interferometer Space Antenna ({\it LISA}). Here we assume that GWs from a MBH binary have been detected by {\it LISA}, and we assess whether or not its electromagnetic (EM) counterpart  can be uniquely identified in archival LSST data as a periodic quasar. We use the binary's properties derived from the {\it LISA} waveform, such as the past evolution of its orbital frequency, its total mass, distance and sky localization, to predict the redshift, magnitude and historical periodicity of the quasar expected in the archival LSST data. We then use Monte Carlo simulations to compute the false alarm probability,
i.e. the number of quasars in the LSST catalogue matching these properties by chance, based on the (extrapolated) quasar luminosity function, the sampling cadence of LSST,  and intrinsic "damped random walk (DRW)" quasar variability. We perform our analysis on four fiducial {\it LISA} binaries, with total masses and redshifts of $(M_{\rm bin}/{\rm M_{\odot}}, z) = (3\times10^5, 0.3)$, $(3\times10^6, 0.3)$, $(10^7, 0.3)$ and $(10^7, 1)$. While DRW noise and aliasing due to LSST's cadence can produce false periodicities by chance, we find that the frequency chirp of the {\it LISA} source during the LSST observations washes out these noise peaks and allows the genuine source to stand out in Lomb-Scargle periodograms. 
We find that all four fiducial binaries yield excellent chances to be uniquely identified, with false alarm probabilities below $10^{-5}$, a week or more before their merger.  This then enables deep follow-up EM observations targeting the individual EM counterparts during their inspiral stage. 
\end{abstract}


\section{Introduction}\label{sec:intro}

Through its upcoming all-sky time domain survey, the Legacy Survey of Space and Time (LSST), the Vera C. Rubin Observatory is expected to observe up to 100 million luminous quasars \cite[][hereafter XH21]{Xin2021}. The optical light curves of the quasars will be archived before the launch of the Laser Interferometer Space Antenna ({\it LISA}), a milli-Hertz gravitational wave (GW) detector sensitive to massive black hole (MBH) binaries with masses of $\approx10^{4-7}{\rm M_{\odot}}$, out to essentially all redshifts  where these MBHs exist \citep{LISA2024}. Assuming that quasars are triggered by galaxy mergers which also correspond to MBH mergers, XH21 find that LSST's quasar catalog will contain $\approx$150 ultra-short period binaries, which will evolve into the {\it LISA} band and subsequently can be detected as GW sources within 15 years after {\it LISA}'s launch. Since these binaries can be identified before LISA's launch, they can serve as ``Verification binaries", in analogy with known compact white dwarf binaries \citep{Burdge+2020}. 

One may imagine the reverse situation, in which a MBH binary's GWs have been detected by {\it LISA}, and ask whether or not an electromagnetic (EM) counterpart of the binary can be identified in archival LSST data.  The EM precursors of {\it LISA} binaries, as discussed in XH21, can be variable quasars with ultra-short periods of $P_{\rm orb} \lesssim {\rm few}$ days. Identifying the EM counterpart of a {\it LISA} binary before merger can facilitate follow-up EM observations for monitoring binary variability during the late stages of GW inspiral and even shortly after merger, in addition to its archival optical variability in LSST. Hydrodynamical simulations have shown that tell-tale EM signatures of the binary can emerge during the last weeks to hours before merger \citep{Farris+2015,Krauth+2023,Dittmann2023,Franchini2024}
and signatures from sudden mass-loss and recoil can build up weeks or months after merger \citep[e.g.][]{Milosavljevic2005,Corrales2010,Rossi2010}. 

Identifying the EM counterparts when we already have both {\it LISA} and LSST data in hand is less challenging than with the LSST light curves alone. The advantages of having {\it LISA} data include the following. (i) {\it LISA} can localize the binary's sky position to within 0.1-1000 deg$^2$ prior to merger, depending on the binary's mass and luminosity distance. The localisation improves as more cycles of GWs are detected. Assuming that {\it LISA} localizes the source to $\sim$10 deg$^2$, and LSST's sky coverage is 18000 deg$^2$, the number of potential candidates in the LSST catalog is reduced by a factor of $>1000$. Given further constraints on the distance and mass of the binary, the number of candidates can be further reduced by 1-2 orders of magnitude, as will be calculated in this work using the extrapolated quasar luminosity function and other quasar properties constrained by {\it LISA} and LSST.
(ii) More importantly, the binary's orbital evolution is driven by GW inspiral during the last years of its merger \citep{Peters1963}, with so-called 'environmental effects' from surrounding stars or gas expected to play a negligible role~\citep[e.g.][]{Haiman2009a,Bortolas+2021}.
As a result, the {\it LISA} data determines the precise orbital frequency evolution ($f$ and $\dot{f}$) of the binary, which depend on its chirp mass and redshift. Without {\it LISA}, identifying $f$ and $\dot{f}$ for a periodic quasar requires much more intricate methods. As a result, without knowing $f$ and $\dot{f}$ to high certainty, identifying a MBH binary in a large sample of variable quasars can yield high false alarm probabilities (see our discussion in \S~\ref{sec:results}).

In this work, we construct mock LSST quasar light curves that consist of two components -- an intrinsic quasar variability component that is characterized by the so-called damped random walk (DRW) process \citep{Kelly2009,MacLeod2010,Kozlowski2010,Ivezic2013}, and an additional periodic modulation arising from the binary nature of the quasar.  For the latter, we adopt a sinusoidal variability for simplicity.  
To the extent that gas co-moving with the BH components contributes to the flux in the LSST band, the relativistic Doppler boost (DB; \citealt{Dorazio+2015}) would produce such sinusoidal variability. Alternatively, hydrodynamical effects during accretion from a circumbinary disk have been shown in numerous simulations to produce periodic variability on timescales commensurate with the orbital period \citep[e.g.,][]{Duffell2020,Zrake+2021,Westernacher-Schneider2022}, and with pulse shapes that become increasingly sinusoid-like as the mass ratio becomes more unequal~\citep{Farris+2014,Duffell2020}. 
Additionally, DRW and aliasing which arises from the LSST cadence can cause periodicity to arise by chance \citep{Vaughan2016}. We use Lomb-Scargle (LS) periodograms to identify the periodicity of our mock LSST light curves, allowing us to calculate the false alarm probability as a result of DRW variability alone. Our goal is to find the false alarm probabilities of identifying EM counterparts in LSST for four fiducial MBH binaries in {\it LISA}. 

This paper is organized as follows. 
In \S~\ref{sec:f_obs}, we derive the time-evolution of the binary's orbital frequency and its uncertainty based on the {\it LISA} data. We then extrapolate the orbital frequency back $\sim$10 years in order to determine the historic binary period in the archival LSST data. 
In \S~\ref{subsec:MockLC}, we compute the expected number $N_Q$ of quasar counterpart candidates for each fiducial {\it LISA} binary, based on the {\it LISA} error boxes and the uncertainties in the binary mass and redshift. We then create mock LSST light curves for each quasar with the combination of the sinusoidal variability and the DRW model, using the expected sampling rate and baseline in LSST's $i$-band observations. 
In \S~\ref{sec:false_alarms}, we quantify the false alarm probabilities arising from DRW variability and aliasing effects, by modeling 100 mock light curves with the DRW model alone for each of the $N_Q\gg 1$ quasars in the {\it LISA} error volume.
We report our results in \S~\ref{sec:results}, where we also discuss a few other studies related to this work and our planned follow-up analysis.
Finally, we summarize our findings and the implications of our results in \S~\ref{sec:summary}. 

\section{orbital frequency of {\it LISA} binaries} \label{sec:f_obs}

\citet[][hereafter M20]{Mangiagli2020} calculate the constraints on the physical properties of a handful of {\it LISA} binaries from their GW waveforms and the expected {\it LISA} noise spectral density. In this work, we adopt the results from M20 on the uncertainties in the chirp mass ($\Delta \mathcal{M}$), luminosity distance ($\Delta d_L$) and sky localization ($\Delta \Omega$) for nine hypothetical {\it LISA} sources with total masses of $M_{\rm bin}=3\times 10^5 {\rm M_{\odot}}, 3\times 10^6 {\rm M_{\odot}}$ and $10^7 {\rm M_{\odot}}$ and redshifts of $z=0.3, 1$ and 3. 

\begin{figure}
    \centering
    \includegraphics[width=\columnwidth]{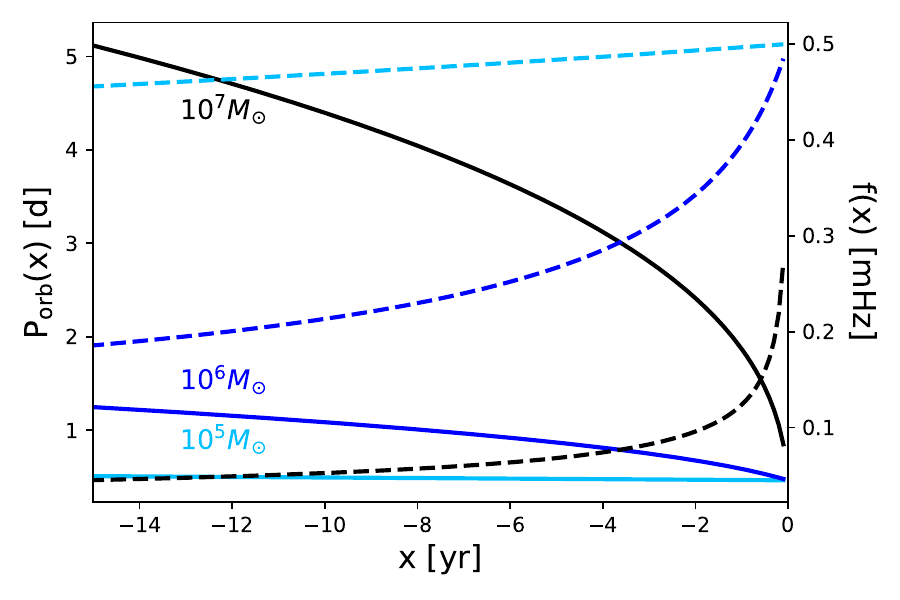}
    \caption{Observed binary orbital period (solid)
    and observed GW frequency (dashed) as a function of time (as shown in Eq.~\ref{eq:f_x}), for binaries of total mass $M_{\rm bin}=10^5 {\rm M_{\odot}}$ (light blue), $10^6 {\rm M_{\odot}}$ (blue), and $10^7 {\rm M_{\odot}}$ (black) at redshift $z=1$ . 
    }
    \label{fig:Porb}
\end{figure}

In this section, we consider an illustrative fiducial binary with $\log{(M_{\rm bin}/{\rm M_{\odot}})}=7$ and redshift $z=1$, and compute its frequency evolution. The rest-frame GW frequency of this binary one week before merging in the {\it LISA} band is approximately $f_r=0.1$ mHz. The corresponding observed (redshifted) binary orbital period is then $P_{\rm orb}=2(1+z)/f_r\sim 0.1$ day. If we assume that the binary orbit is circular and the binary inspiral is driven purely by GW emission during the last few years before merger, the change in rest-frame GW frequency  (or frequency chirp) can be described as, 
\begin{equation} \label{eq:f_dot}
    \dot{f_r} = \frac{96 \pi^{8/3}G^{5/3}}{5c^5} \mathcal{M}_c^{5/3} f_r^{11/3}, 
\end{equation}
where $\mathcal{M}_c$ = $[q/(1+q)^2]^{3/5}M_{\rm bin}$ is the binary's chirp mass and $q\equiv M_2/M_1\leq 1$ is its mass ratio. Note that throughout this paper, we assume $q=0.1$, a commonly expected mass ratio of {\it LISA} MBH binaries \citep{Katz2020}. The frequency chirp is more rapid for more massive binaries, and for binaries closer to merger. This form of chirping not only dictates the binary's GW frequency evolution in the {\it LISA} band, but also predicts the evolution of the orbital period in the archival LSST data. Specifically, integrating Eq.~\ref{eq:f_dot} backward over time gives us the rest-frame GW frequency as a function of the (rest-frame) look-back time $x_r$. Defining $x_r=0$ as a characteristic reference time of the {\it LISA} detection, we find
\begin{equation} \label{eq:f_x_r}
    f_r(x_r) = \left[{f_r(0)}^{-8/3} - \frac{256 \pi^{8/3}G^{5/3}\mathcal{M}_c^{5/3}}{5c^5} x_r \right]^{-3/8},
\end{equation}
or, in terms of observed (Earth-frame) time $x$ and frequency $f$,
\begin{equation} \label{eq:f_x}
    f(x) = \left\{{f(0)}^{-8/3} - \frac{256 \pi^{8/3}G^{5/3}[(1+z)\mathcal{M}_c]^{5/3}}{5c^5} x \right\}^{-3/8}.
\end{equation}
Let $f(0)$ be a characteristic frequency at which the binary is detected by {\it LISA}, which we assume to be $\sim 10^{-4}$ Hz (i.e. approximately a week prior to merger, as noted above, for the fiducial binary). The same $f(0)$ corresponds to $\sim3$ weeks and $\sim2.7$ years prior to merger for $M_{\rm bin}=3\times10^6 {\rm M_{\odot}}$ and $3\times10^5 {\rm M_{\odot}}$, respectively, at redshift $z=1$. In Fig.~\ref{fig:Porb}, we show $f(x)$ (dashed curves) and $P_{\rm orb}(x)$ (solid curves) for binaries with three different masses at redshift $z=1$, $M_{\rm bin}=10^5{\rm M_{\odot}}$ (cyan), 
$10^6{\rm M_{\odot}}$ (blue) and $10^7{\rm M_{\odot}}$ (black). If we extrapolate the time back $10$ years, the periodically varying quasar
corresponding to this binary will exhibit the period evolution shown in this figure. 
For ``Verification binaries", ten years prior to the {\it LISA} detection, the (observed) orbital periods range between $P_{\rm orb}\approx$ 0.5 -- 5 days. 

\subsection{Uncertainty of the historical orbital period} \label{sec:f_err} 

The uncertainty in the observed frequency, $\delta{f}$, 
can be estimated from the binary parameters constrained by the GWs. In the simplest case of a circular, non-spinning binary, the frequency $f(x)$ at the look-back time $x$ has two sources of uncertainty, from the instantaneous frequency $f(0)$ measured in the {\it LISA} band, and from the inferred (redshifted) chirp mass $\mathcal{M}_z=(1+z)\mathcal{M}_c$ that dictates the chirping rate (see Eq.~\ref{eq:f_x}).  

A rough estimate of the fractional error on $f(0)$ can be obtained by considering the uncertainty of the position of the peak in the Fourier transform of a sinusoidal strain signal $h(t)$ over a finite observational window.  If the number of GW cycles detected around the frequency $f(0)$ is $N_{\rm cyc}$,
then the Delta-function in Fourier space (for an infinite time-series data) is broadened to have a width of $f(0)/N_{\rm cyc}$ around $f(0)$. The positioning error of this broadened peak (e.g. obtained from matched filtering) should further scale as 1/SNR, where SNR is the instantaneous signal-to-noise ratio at the time of the ${\it LISA}$ observation.   Thus, neglecting time-sampling and other sources of error, we expect $\delta f(0)/f(0)\sim 1/({\rm SNR}\times N_{\rm cyc})$.  

The SNR values for {\it LISA} binaries can be estimated using the observed GW strain and the detector sensitivity. For the binaries presented in M20, SNR roughly goes from $\sim1-10^3$. The number of GW cycles detected depends on the total phase ($\Phi$) as $N_{\rm cyc}=\Phi/2\pi$, where
\begin{equation} \label{eq:phase}
    \Phi = 2\pi \int_{f_{\rm min}}^{f_{\rm max}} f/\dot{f} df.
\end{equation}
It follows from Eq.~\ref{eq:f_dot} and Eq.~\ref{eq:phase} that $N_{\rm cyc}$ can be expressed as 
\begin{equation} \label{eq:n_cyc}
    N_{\rm cyc} = -\frac{c^5}{32\pi^{8/3}G^{5/3}{\mathcal{M}_z}^{5/3}} f^{-5/3} \bigg\rvert_{f_{\rm min}}^{f_{\rm max}}. 
\end{equation}
As more cycles of GWs are detected, the error in $f$ will become smaller. 
For example, for our fiducial binary, $N_{\rm cyc} \approx 85$ from $f_{\rm min}=0.05$ mHz to $f_{\rm max}=0.1$ mHz. 
Therefore, $\delta{f_0}\equiv \delta{f(0)}=f(0)/({\rm SNR} \times N_{\rm cyc})=10^{-8}$ Hz for our fiducial binary a week prior to merger (SNR $\approx 100$).

Next, we estimate this uncertainty for LSST light curves, $x$ years prior to the {\it LISA} detection. For this, we use error propagation, 
\begin{align}\label{eq:err_prop}
\begin{split}
    \delta{f}^2 = \bigg( \frac{\partial{f}}{\partial{f_0}}\bigg)^2 \delta{f_0}^2 + \bigg( \frac{\partial{f}}{\partial{\mathcal{M}_z}}\bigg)^2  \delta{\mathcal{M}_z}^2 & + \bigg(\frac{\partial{f}}{\partial{x}}\bigg)^2 \delta{x}^2 \\ 
    & + COV(f_0,\mathcal{M}_z,x)~{\rm terms}.
\end{split}
\end{align}
Assuming that there is no uncertainty on the time, $\delta{x}$=0, and the covariances between $COV(f_0,\mathcal{M}_z,x)$ are zero, it follows that $\delta{f}$ can be expressed as
\begin{equation} \label{eq:df_simp}
    \delta{f} = \sqrt{\bigg( \frac{\partial{f}}{\partial{f_0}}\bigg)^2 \delta{f_0}^2 + \bigg( \frac{\partial{f}}{\partial{\mathcal{M}_z}}\bigg)^2 \delta{\mathcal{M}_z}^2}, 
\end{equation}
where $f_0=10^{-4}$ Hz and $\delta{f_0}\sim 10^{-9}$. 
Combining with Eq.~\ref{eq:f_dot}, we find that the partial derivatives can be expressed as,
\begin{equation} \label{eq:part_f}
    \frac{\partial{f}}{\partial{f_0}} = [1 - \mathcal{C}\mathcal{M}_z^{5/3}{f_0}^{8/3}x]^{-11/8} 
    {\rm \ \ and }
\end{equation}
\begin{equation}\label{eq:part_M}
    \frac{\partial{f}}{\partial{\mathcal{M}_z}} 
    = \frac{5}{8} \mathcal{C}\mathcal{M}_z^{2/3}x[{f_0}^{-8/3}-\mathcal{C}\mathcal{M}_z^{5/3}x]^{-11/8},
\end{equation}
where we have introduced the constant $\mathcal{C} \equiv 256\pi^{8/3}G^{5/3}/5c^5$. From Eq. \ref{eq:df_simp} to Eq. \ref{eq:part_M}, we find that $\delta{f}/f\approx 0.06$ when $x=-10$ yr. 
Alternatively, since we know that the chirp mass term dominates the frequency evolution, we can directly write $\delta{f}/f \propto (5/8) \delta{\mathcal{M}_z}/\mathcal{M}_z$, from Eq.~\ref{eq:f_x}. According to Fig.~4 in M20, for a binary with $M_{\rm bin}=10^7{\rm M_{\odot}}$ and $z=1$ a week before merger, the uncertainty on the redshifted chirp mass is $\delta{\mathcal{M}_z}\approx 0.1\mathcal{M}_z$.\footnote{Note that Fig.~4 in M20 refers to chirp mass, but we have confirmed that the figure, in fact, shows relative errors on the {\em redshifted} chirp mass (Mangiagli, private communication).} 
It follows that $\delta{f}/f = 6.2\%$, which is consistent with our calculation from the error propagation above. The specific value of $\delta{f}$ changes with time since $f$ changes with time. But in general, $\delta{f}$ gives the allowed range of the frequencies in which the peak of the periodicity of the LSST light curve is expected to occur, indicating that this light curve can be a true counterpart of the {\it LISA} binary. M20 also shows that $\delta{\mathcal{M}_z}$ is smaller for less massive black hole binaries -- $\delta{\mathcal{M}_z}/\mathcal{M}_z=0.07$ and $0.02$ for $M_{\rm bin}=3\times10^6{\rm M_{\odot}}$ and $M_{\rm bin}=3\times10^5{\rm M_{\odot}}$ BHs, respectively. Therefore, the uncertainties of $\delta{f}/f$ for fiducial models with the other two binary masses are $4.4\%$ and $1.3\%$, respectively. 

\section{Mock LSST light-curves} \label{subsec:MockLC}

Out of the nine {\it LISA} sources reported in M20, we find that only four may produce apparent i-band magnitudes above LSST's detection limit of $m_i=26$. The magnitude $m_i$ for each binary can be estimated from its mass and redshift as \citep{Haiman2009a},
\begin{equation} \label{eq:mi}
        m_i = 26 + 2.5 \: {\rm log} \: \bigg[ \bigg( \frac{{\rm f}_{\rm Edd}}{0.3}\bigg)^{-1} \: \bigg( \frac{M_{\rm bin}}{3\times 10^7 {\rm M}_{\odot}}\bigg)^{-1} \bigg( \frac{d_L(z)}{d_L(z=2)}\bigg)^2 \: \bigg].
\end{equation}
This relationship is based on the mean quasar spectral energy distribution and adopts a fiducial bolometric quasar luminosity corresponding to an Eddington ratio $f_{\rm Edd} = L/L_{\rm Edd}=0.3$.
The same relationship was also used in XH21. 
Using Eq.~\ref{eq:mi}, we find that the four {\it LISA} binaries with log($M_{\rm bin}$/${\rm M_{\odot}}$), $z$) = (5.5,0.3), (6.5,0.3), (7,0.3) and (7,1) have $m_i\leq 26$. These are the four binaries that we focus on in this work. We assign model numbers and summarize some of their properties in Table~\ref{tab:model_summary}. In particular, we have already discussed our method to find their historical orbital periods ($P_{\rm orb}$) in \S~\ref{sec:f_obs} above.

In this section, we generate mock light curves for each of these four binaries. 
We first determine the number of quasars, $N_{\rm Q}$, expected to be catalogued by LSST around each of the four $(M_{\rm bin},z)$ values (\S~\ref{subsec:n_qso}). This determines the number of mock light-curves needed to reliably compute a false-alarm probability. We then discuss (\S~\ref{subsec:sin_drw}) the ingredients of the mock light curves, including sinusoidal variability from binary accretion,
intrinsic stochastic variability (damped random walk), and photometric errors (assumed to be Gaussian with an r.m.s. of 1 magnitude). We choose a 10-yr baseline for all light curves. Models (I) and (IV) have magnitudes fainter than the single exposure limit of LSST ($m_i$=24 mag), so they will require co-adding multiple observations. Assuming that the
flux limit improves with the number $N_{\rm exp}$ of co-added exposures as $1/\sqrt{N_{\rm exp}}$, we find that making these sources detectable
would require approximately 40 and 13 co-added LSST data points, corresponding to $\sim 120$ and 40 days, respectively.

\begin{table} 
    \centering
    \begin{tabular}{ |c|c|c|c|c|c| } 
         Model &  (M$_{\rm bin}/{\rm M_\odot}$,z) & $m_i$ [mag] & P$_{\rm orb}$ [d]&  $\Delta \Omega$ [${\rm deg}^2$] & $N_{\rm Q}$  \\ \hline
         (I) & (3$\times 10^5$, 0.3) & 26.0 & 0.45& 6 & 1,014  \\ \hline
         (II) & (3$\times 10^6$, 0.3)& 23.5 & 1.4 &  40 & 5,918 \\ \hline
         (III) & ($10^7$, 0.3) & 22.1 & 3.2 & 200 & 11,054 \\ \hline
         (IV) & ($10^7$, 1) & 25.3 &4.7 & 6,000 & 773,807 \\
    \end{tabular}
    \caption{Properties of the four fiducial {\it LISA} binaries and their light-curves. The columns correspond to (1) model number, (2) total binary mass and redshift, (3)
    apparent i-magnitude, 
    (4) initial orbital period (10 years prior to the {\it LISA} detection), (5) the uncertainty in {\it LISA} sky localization one week before merger, 
    and (6) number of quasars from the extrapolated quasar luminosity function within {\it LISA}'s sky localization error and inferred mass and redshift range.  }
    \label{tab:model_summary}
\end{table}

\subsection{Number of quasars} \label{subsec:n_qso}

\begin{figure}
    \centering
    \includegraphics[width=\columnwidth]{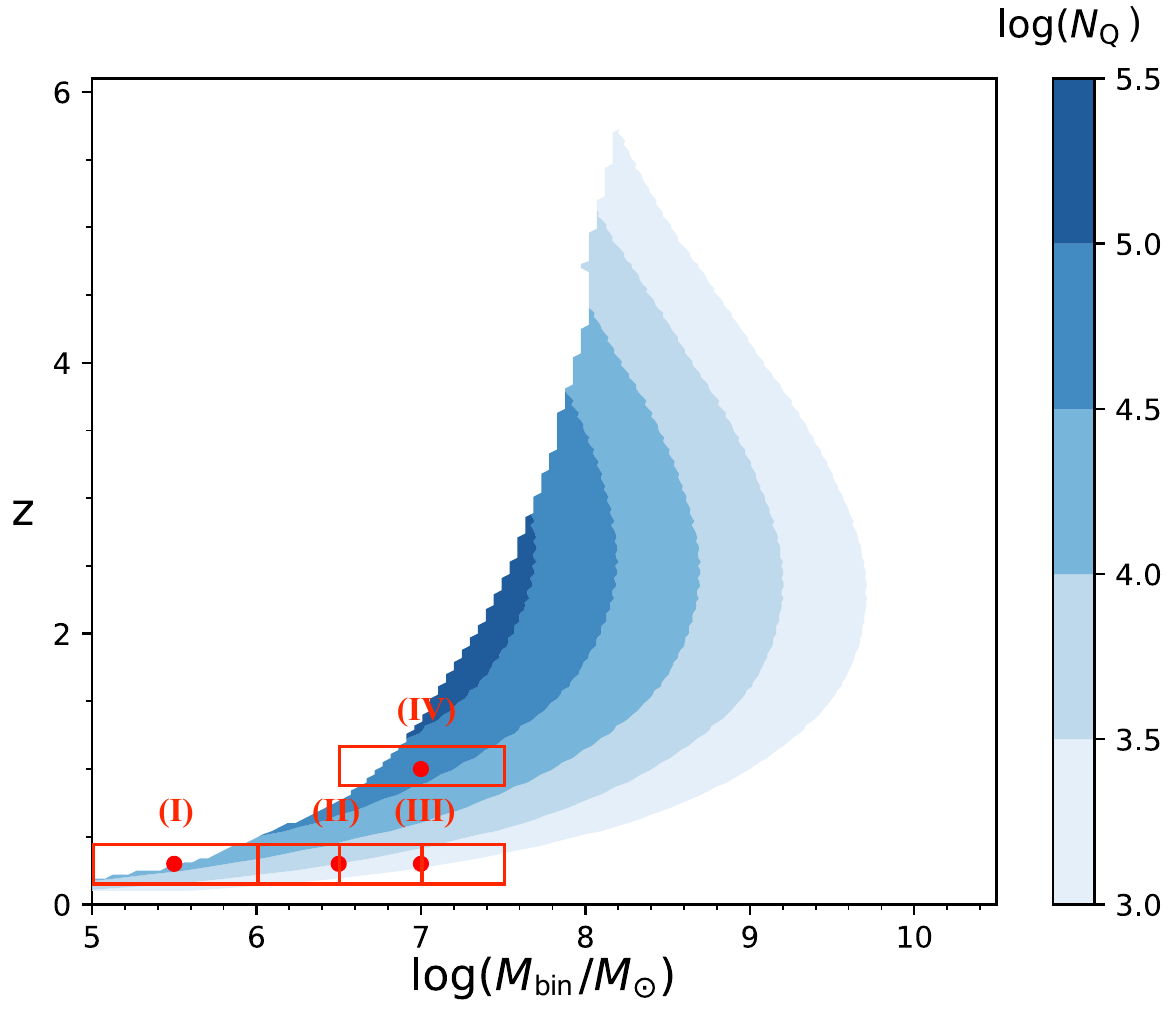}
    \caption{The number of LSST quasars ($N_{\rm Q}$) predicted from the QLF, following the results in XH21. This contour shows a total of $\sim$100 million quasars. The masses and redshifts of the four fiducial {\it LISA} binaries are colored with red dots, with their corresponding models numbers in the brackets. The error boxes have widths of 1 and heights of 0.2, coming from the uncertainties in BH mass and photometric redshift, which yield $\Delta\log(M_{\rm bin}/{\rm M_{\odot}})=+/- 0.5$ and $\Delta z = +/- 0.1$, respectively. 
    }
    \label{fig:N-color}
\end{figure}

According to XH21, LSST will discover and measure the light-curves of up to 100 million quasars. They identify the number of quasars ($N_{\rm Q}$) in LSST as a function of mass and redshift by extrapolating empirical quasar luminosity functions \cite[QLFs,][]{Kulkarni2019}. In practice, the QLFs are determined in discrete redshift bins, and fit with parametric functions; X21 interpolate and extrapolate the four QLF parameters as smooth functions of redshift from $z=0$~to~6. We summarize the findings of XH21 in Fig.~\ref{fig:N-color}, where $N_{\rm Q}$ is the number of quasars expected in LSST projected on the $M_{\rm bin}-z$ plane. Between $\log(M_{\rm bin}/{\rm M_\odot})=5-9.5$ and redshifts up to $z=6$, there are a total of $\sim$100 million quasars. Note that the extrapolated QLF we adopt in XH21 is not ideal for redshifts much larger than 6, especially for faint quasars, since the QLFs have not been designed to accommodate the $z>6$ range. In fact, there is no observational data coverage in the adopted QLFs in XH21 (also see Fig. 3 in XH21). There have been $\sim$300 known luminous quasars at $z>6$ \citep[e.g.][]{Fan2006,Willott2010,Yang2019,Banados2023}, and this population is expected to extend with JWST observations. Our extrapolated QLF predicts around 900 LSST quasars in the redshift range $z=6-8$, with magnitudes limit $m_i=26$. However, the objects we consider (i.e. the ``Verification binaries") are mostly distributed at $z\lesssim 3$. 
The red points represent the four fiducial {\it LISA} binaries, denoted with their model numbers, and the contour scale on the right is in units of number of quasars in bins of width $\delta\log M_{\rm bin} = 0.01$ and $\delta z = 0.01$.

The red boxes shown in Fig.~\ref{fig:N-color} represent crude estimates of the mass and redshift bins one may search for a LISA counterpart.  The uncertainty on the mass  $M_{\rm bin}$  
of the quasar is assumed to be an order of magnitude ($\Delta\log(M_{\rm bin})=+/- 0.5$), capturing the range of Eddington ratios for bright quasars \citep{Kollmeier+2006,Hopkins2009,Kochanek2012}.
The redshift uncertainties represent photometric redshifts errors. 
These errors can be as bad as $\Delta z = +/- 0.1$ for 90\% of the quasars, including faint quasars with i-mag $\lesssim 25$~\citep{LSSTScienceCollaboration2009}. 

Finally, Fig.~\ref{fig:N-color} shows the number of quasars over the whole sky, while the {\it LISA} sky localization helps to narrow the search down to a smaller area. Generally, the sky position errors ($\Delta \Omega$) not only depend on the mass and luminosity distances of the source, they also become smaller with time or as more GW cycles are detected. To be consistent with the exercise in \S~\ref{sec:f_obs}, we adopt the values of $\Delta \Omega$ a week before merger (from Fig. 7 in M20). Although better sky localization can be achieved closer to (or past) the merger, ideally we want the time of detection to be at least one week prior to merger to facilitate follow-up EM observations before the merger occurs. 
We thus rescale $N_{\rm Q}$ by $\Delta \Omega$/(18000 deg$^2)$, where 18000 deg$^2$ is LSST's sky coverage. 
We summarize the final values of $N_{\rm Q}$ within the mass, redshift, and sky location range for each model in Table~\ref{tab:model_summary}.  As the table shows, the number of quasars is still large, and a unique counterpart cannot be identified simply based on its sky position, redshift, and inferred mass.  We next turn to periodicity. 

\subsection{Sinusoidal variability and damped random walk} \label{subsec:sin_drw}
 \begin{figure} \label{fig:mock_lc}
    \centering
    \includegraphics[width=\columnwidth]{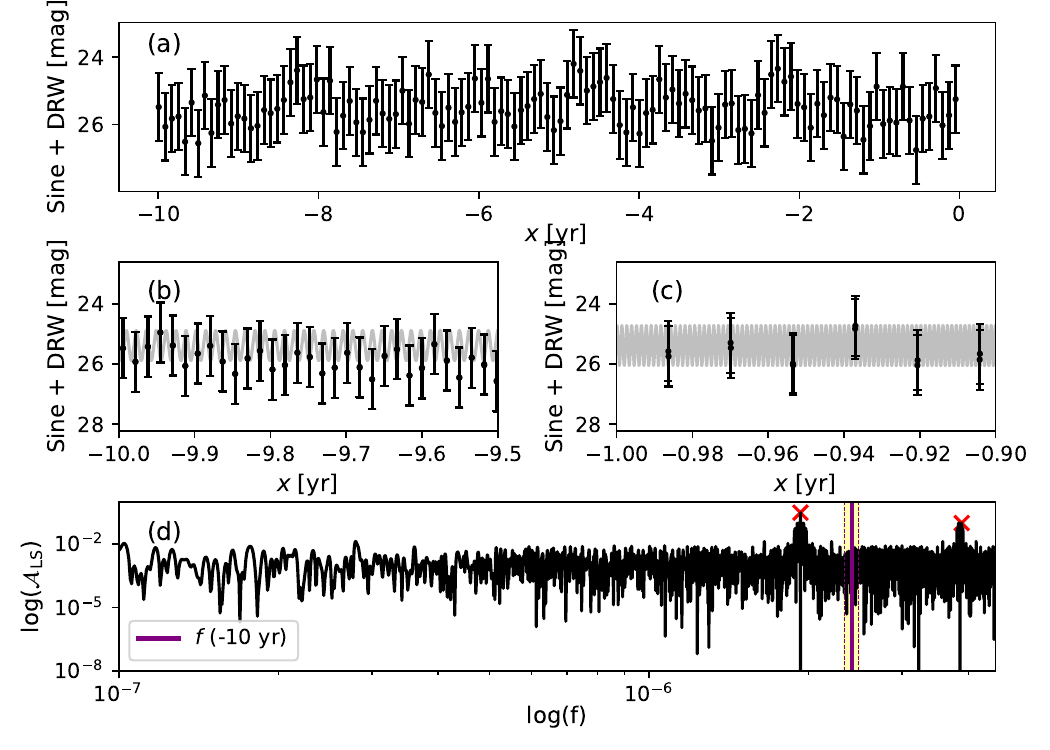}
    \caption{This figure shows (a) an example mock light curve, plotted for every 10th LSST data point, which incorporates periodic sinusoidal variability, red-noise damped-random walk variability, and photometric errors of 1 mag for a MBH binary with $M_{\rm bin}=10^7{\rm M_{\odot}}$ and $z=1$. In panels (b) and (c), we show the light curve at the beginning (first 0.5 years of observation) and near the end (from -1 to -0.9 years) of the LSST observations, with the sine wave superimposed. These segments of the light curve demonstrate that the orbital period increases, specifically from 4.7 days at the beginning of the LSST observations to 1 day towards the end.  Panel (d) shows the Lomb-Scargle periodogram of the entire light curve, which shows no significant peak within $f+/-\delta f$ (yellow shaded region between the purple lines) due to the rapid chirping of the binary. The solid purple line represents the observed frequency $f({\rm -10~yr})$. We do, however, observe two peaks (red crosses) outside the bin due to pure noise.  }
    \label{fig:mocklc_pg}
\end{figure}

In Fig.~\ref{fig:mocklc_pg}, we show an example mock light curve for a MBH binary with $M_{\rm bin} = 10^7 {\rm M_{\odot}}$ and $z=1$ (model IV). Our mock light curves 
include a periodic sinusoidal modulation and intrinsic DRW red noise, as discussed above.
The baseline for this light curve is 10~years.
We avoid showing the entire 10-yr light curve, with a pair of data points every 6 days \citep{SurveyCadence}, since we cannot visually extract information from such a densely distributed time series. Instead in panel (a), we show the light-curve down-sampled at every 10$^{\rm th}$ observation. In panels (b) and (c), we show the the light curve in the first 0.5 years of the LSST observations, and again near the end, from -1 to -0.9 years, to demonstrate the change in orbital period. The photometric errors are fixed to 1 mag. 

For illustration, 
we adopt a sinusoidal variability for the optical light curves caused by relativistic Doppler boost \citep{Dorazio+2015}.
As mentioned above, hydrodynamical simulations have shown more complex pulse shapes of variability such as saw-toothed or spikey light curves \citep{Zrake+2021,Westernacher-Schneider2022,Cocchiararo2024} 
whose shapes depend on the eccentricity and mass ratio. Here we adopt the sinusoidal variability for simplicity, as shown superimposed in the middle panels of Fig.~\ref{fig:mocklc_pg}.
If it arises from Doppler boost, the sinusoidal fluctuation, in the limit of small amplitudes, takes the form of 
$m_i+ A(x){\rm sin}[2\pi f(x) \cdot x]$, where $m_i$ is the average magnitude (25.4 mag as listed in Table~\ref{tab:model_summary}) 
of the light curve, and 
$f(x)$ is the frequency evolution given in Eq.~\ref{eq:f_x}.
The amplitude of the sine wave at -10 years is 0.5 mag which can be consistent with both the amplitude of Doppler boost as well as hydrodynamical variability. The amplitude $A(x)$ then increases with time as the orbital velocity $v/c$ increases. From panel (b) to panel~(c) in Fig.~\ref{fig:mocklc_pg}, the amplitude increases from 0.5 to 1.07 mag. We also see the effect of frequency chirping - the observed orbital period is initially 4.7~d in panel (b), and it evolves to be approximately 1 day in panel (c), a year before the end of LSST observation. 

We also implement the DRW model for the stochastic noise/variability of quasars. DRW is characterized by an amplitude and a timescale that depend on the MBH's total mass, redshift, absolute i-magnitude and the rest-frame wavelength. 
We use the best-fit DRW parameters below~\citep[see Eq.~7 in][]{MacLeod2010}
\begin{equation} \label{eq:drw}
    {\rm log}~q = a + b~{\rm log}~\Bigg(\frac{\lambda_{\rm RF}}{4000 \angstrom} \Bigg) + c~(M_i+23) + d~{\rm log}~\Bigg(\frac{M_{\rm bin}}{10^9 {\rm M_{\odot}}} \Bigg) + e~{\rm log}~(1+z).
\end{equation}
For each of the four MBH binaries, we calculate $q$, which represents the DRW parameter, the characteristic amplitude SF$_{\infty}$ or the rest-frame variability timescale $\tau$, with the best-fit coefficients $(a, b, c, d, e) = ($-0.51, -0.479, 0.131, 0.18, 0.0$)$ or $($2.4, 0.17, 0.03, 0.21, 0$)$, respectively.
We assume $\lambda_{\rm RF}=7000/(1+z)$ \AA \ for the LSST $i$-band in our calculations. For the four quasars in this study, the typical DRW variability timescale and characteristic amplitude are $\tau \sim 100$ days and ${\rm SF}_{\infty}=0.3$. 

Finally, we use the Lomb-Scargle (LS) periodogram, a common Fourier technique to find periodicities in unevenly sampled light curves. {\it A unique period cannot be picked out in the periodogram because the frequency is chirping significantly during the LSST  observations}, as shown in the bottom panel~(d) of Fig.~\ref{fig:mocklc_pg}. As a reference, we show the initial orbital frequency 10 years before merger ($f_0$; purple dotted line) and the allowed range for the location of the peak (in-between the thin purple lines) according to our estimation in \S~\ref{sec:f_err} -- recall $\delta f/f=6.2\%$ for the $10^7{\rm M_{\odot}}$ MBH binary. The red crosses represent peaks that arise from pure noise; see the next section for a detailed discussion. 

\begin{figure}
\centering
\includegraphics[width=\columnwidth]{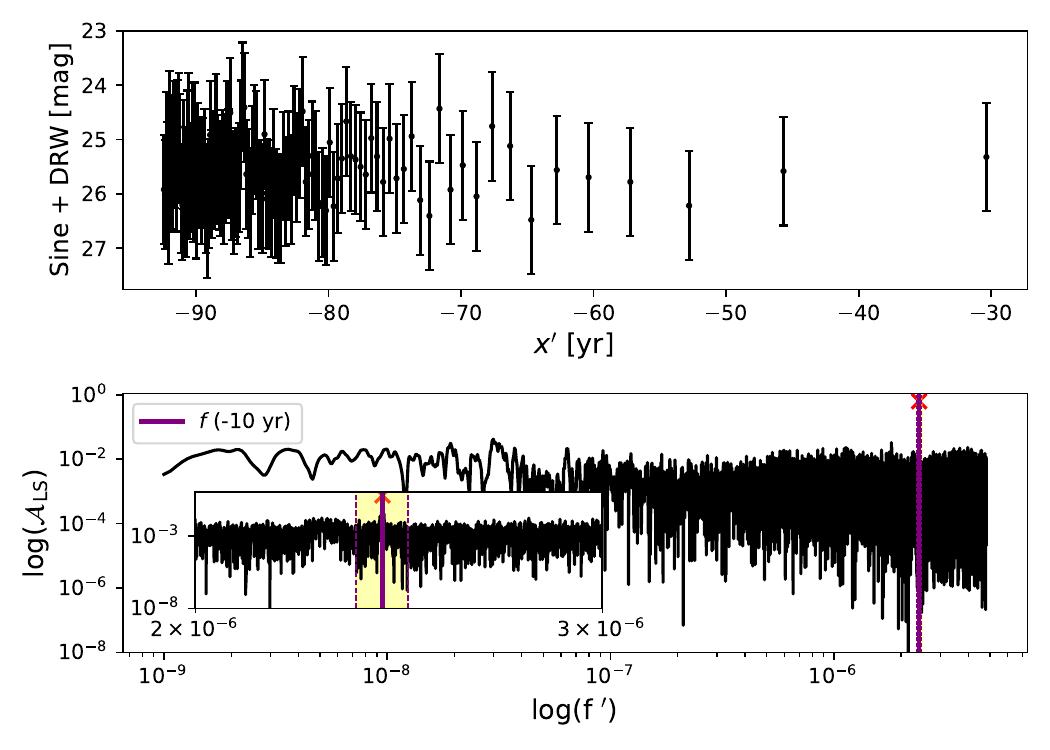}
\caption{{\it Top:} same light curve as Fig.~\ref{fig:mocklc_pg}, plotted for every 10$^{\rm th}$ data point, but with time rescaled as $x^{\prime}=[f(x)/f({\rm -10~yr})] \ x$, to remove the effect of chirping. The stretched out light curve has a constant period at $f^\prime({\rm -10~yr})^{-1}$, and the same number of cycles as the light curve in Fig.~\ref{fig:mocklc_pg}. {\it Bottom:} the LS periodogram of this stretched light curve, which now shows a prominent peak at $f^\prime({\rm -10~yr})$ with no noise-peaks in the allowed range (see the zoom-in insert). }
\label{fig:lc_xprime}
\end{figure}

Given that we know the exact form of frequency chirping of the binary \citep[assuming that the light curves follow this chirp, which appears justified by hydrodynamical simulations of inspiraling binaries;][]{Tang+2018,Krauth+2023}, 
we can remove the effect of chirping by re-scaling the time axis, $x$, by $x^{\prime} \equiv [f(x)/f({\rm -10~yr})] \ x$ such that the new sine wave in the form of $A(x^{\prime}) \ {\rm sin}(\phi^{\prime} \cdot x^{\prime}) + m_i$, where $\phi^{\prime}=2\pi f({\rm -10~yr})$, will have a fixed periodicity of $f^\prime({\rm -10~yr})$. As a result, the light curve is ``stretched out", see top panel of Fig.~\ref{fig:lc_xprime} (again plotted for every 10$^{\rm th}$ observation for clarity). In the bottom panel, we show that we are able to uniquely identify the true periodicity of this light curve, with no noise-peaks appearing by chance within the allowed width of the peak (see the zoom-in inset near the peak). 

This exercise illustrates the fact that it is challenging to use an LSST light curve alone, without any {\it LISA} information, to find a unique periodicity for an ultra-compact 
MBH binary. This is because it is hard to obtain precise BH mass with LSST alone due to the uncertain Eddington fraction and photometric redshift. We need {\it LISA} GW data to give us the exact form of frequency evolution of the light curves to perform this analysis. 
Additionally, in the example we show in Fig.~\ref{fig:lc_xprime}, we have the most ideal case where the BH binary is the most massive out of the four fiducial binaries. Its orbital frequency, $f({\rm -10~yr})$, is lower than the sampling frequency.
False periodicities could arise from pure noise if the orbital frequency is similar to/above the sampling frequency. In the following section, we discuss the possibility of false peaks arising from stochastic DRW variability and from aliasing effects. 

\section{False alarm probability from DRW and aliasing} \label{sec:false_alarms}

Numerous work have characterized the stochastic variability of luminous quasars with the DRW model \citep{Kozlowski2010,MacLeod2010,Andrae2013}. \citet{Vaughan2016}
has quantified how often fake periodicities in quasar light curves can arise by chance due to DRW variability. We here similarly evaluate the likelihood that peaks in the LS periodograms result from DRW and LSST sampling rate alone. 
The characteristic time-scales of the DRW variability of binaries in this study are typically of order 100 days. It follows that DRW should cause a frequency break at $f \approx 10^{-7}$ Hz on the periodogram, above which the amplitude of the DRW decreases as $f^{-2}$, although the amplitude drop is less steep for variability on shorter timescales associated with lower BH masses (and luminosity) \citep{Caplar2017}, which are the objects most relevant in this study. Note that existing studies on the stochastic variability of quasars only consider higher-mass BHs (e.g. \citealt{Caplar2017} which investigates the variability on a range of timescales, including those from lower-mass BHs, only extrapolate down to $\log(M_{\rm BH}/{\rm M_{\odot}})=7.5$). 

\begin{figure}
    \centering
    \includegraphics[width=\columnwidth]{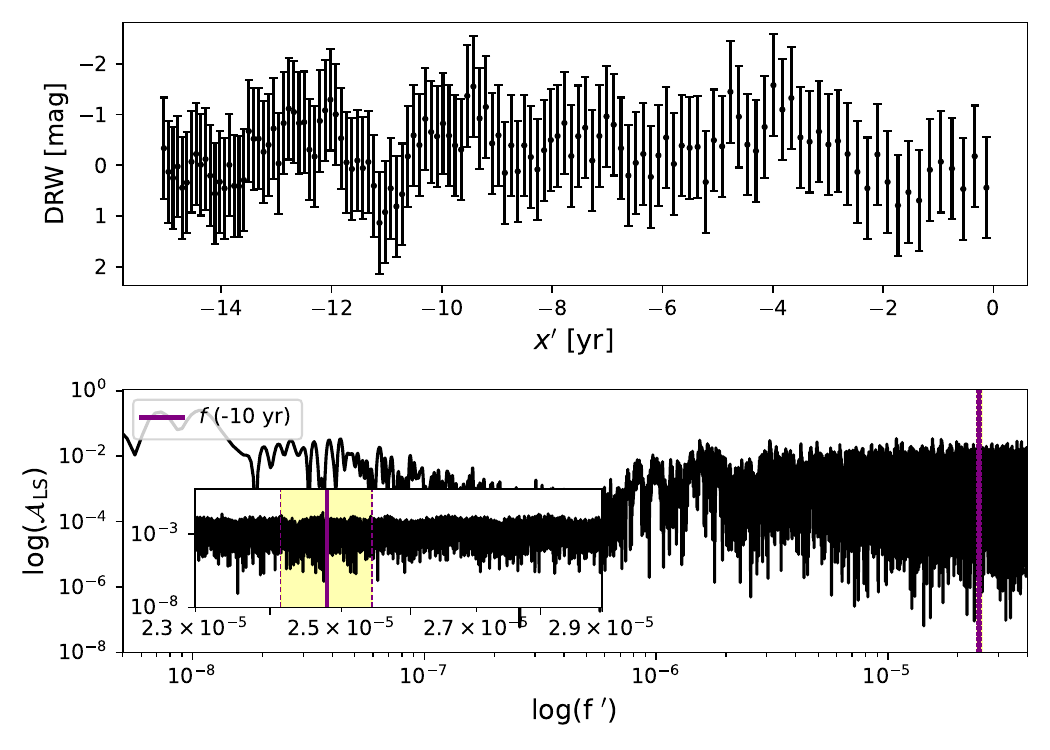}
    \caption{An example of a pure damped random walk light curve (top), plotted for every 10$^{\rm th}$ data point, for a MBH binary with $M_{\rm bin}=3\times 10^5 {\rm M_{\odot}}$ and $z=0.3$, sampled at LSST cadence and its LS periodogram (bottom). We see no peak in its periodogram. }
    \label{fig:drw_pg}
\end{figure}

\begin{figure}
    \centering
    \includegraphics[width=\columnwidth]{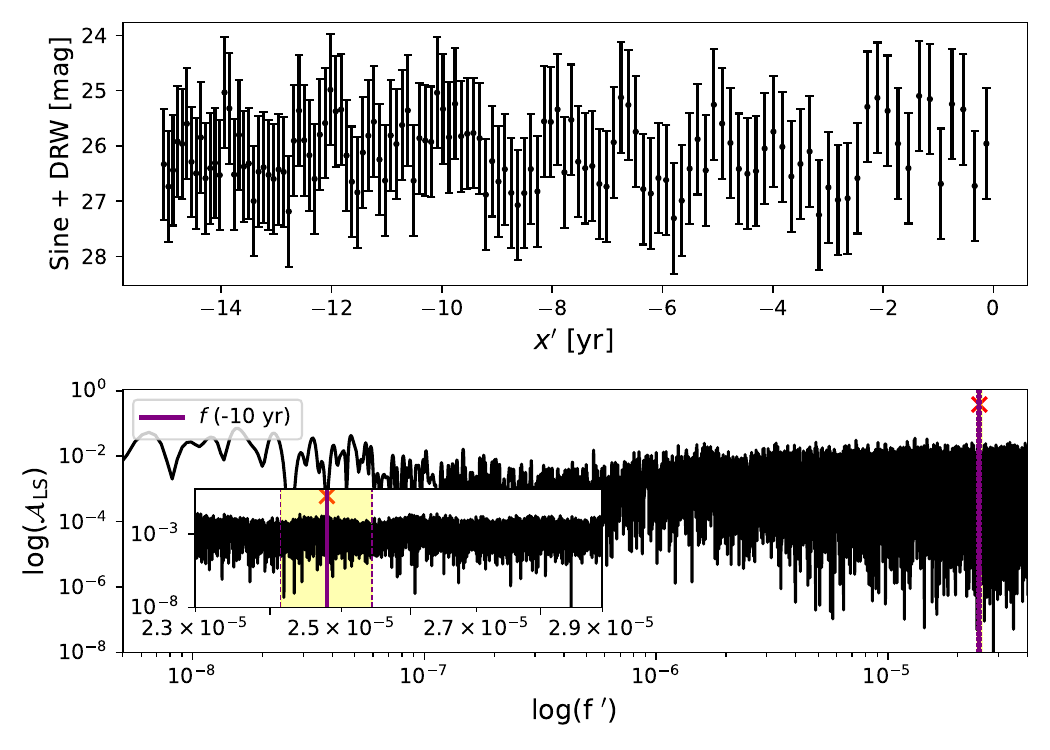}
    \caption{Light curve (top) for the same model as Fig.~\ref{fig:drw_pg}, plotted for every 10$^{\rm th}$ data point, but with a genuine sinusoidal signal injected. Comparing this LS periodogram (bottom panel) with that in Fig.~\ref{fig:drw_pg}, a true peak arises at the expected frequency when the sinusoid is present.  } 
    \label{fig:lc_pg}
\end{figure}

However, generally, DRW can still cause peaks at higher frequencies, although it is harder to mimic the higher amplitude peaks produced by a sinusoidal signal. In Fig.~\ref{fig:drw_pg}, we show an example of a pure DRW light curve for a quasar with mass $M_{\rm bin}=3\times 10^5 {\rm M_{\odot}}$ and redshift $z=0.3$, in the ``stretched" time coordinate (top panel; every 10$^{\rm th}$ data point of the light curve are shown). The bottom panel shows the LS periodogram of this light curve, in which we see a (not so strong) amplitude decrease roughly corresponding to the $1/\tau \sim 10^{-7}$ drop-off from DRW.
We also observe no peak in the $f$(-10 yr)-bin, which indicates that (this particular realization of) DRW light curve does not accidentally produce a false peak at the allowed frequency bin. More importantly, any aliasing peaks that would arise near the $f$(-10 yr)-bin, when the light curve is evenly sampled, have vanished due to the stretching of the time coordinate which continuously widens the sampling cadence in $x^{\prime}$. We discuss the aliasing effect in detail in the next section. We compare Fig.~\ref{fig:drw_pg} with Fig.~\ref{fig:lc_pg}, where the sinusoidal model is injected. Here the periodogram clearly recovers the true frequency of the binary. Fig.~\ref{fig:lc_pg} can also be compared with the Fig.~\ref{fig:lc_xprime}, where the light curve was shown for a different (more massive) binary, in which case the frequency chirps faster and the periodogram peaks at a lower frequency. Generally, it is unlikely that DRW can reproduce the true periodicity of the binary for more than 10 periods. In the archival LSST dataset, there will be approximately a few thousand periods for the light curves of our fiducial binaries, assuming a 10-yr LSST baseline.

To compute the probability that a false peak arises by chance in the $f({\rm -10~yr})\pm\delta f/2$ frequency bin, we perform 100$\times N_{Q}$ Monte Carlo (MC) simulations of the DRW light curves for each binary. For example, for model (I), we have generated 1,014 binary light curves using the Sine + DRW model and 101,400 light curves of the pure DRW case. For each Sine + DRW light curve, we record the maximum height among any local peaks within the $f({\rm -10~yr})$ bin.  We end up with a Gaussian-looking distribution for these peak-heights, $\mathit{A}_{\rm LS}$, with true periodicity. 
For each DRW-only light curve, we record any peak of significance within the $f({\rm -10~yr})$ bin, meaning that the peak height is at least 10\% of the mean height of the ``True peaks". This choice is safe because the distribution in the ``True peaks" varies by no more than 45\% of its average. 
In the ``worst" case of the binary with $3\times 10^5 {\rm M_{\odot}}$ and $z=0.3$, the peak height varies by 45\%, whereas in the other three fiducial models, it typically varies within 30\%. Therefore, any peak from pure noise that is 10\% lower in amplitude than the true peak can be safely attributed as such.
With this criterion, we do not observe any false peaks produced by any realizations of the pure (stretched) DRW light curves, i.e. the distributions of the false peaks are delta-functions at $\mathit{A}_{\rm LS}=0$. 
Therefore, the false alarm probabilities (FAPs) are less than $1/(100\times N_Q)$, which correspond to FAP $\lesssim10^{-5}$, FAP $\lesssim10^{-6}$, FAP $\lesssim10^{-7}$ and FAP $\lesssim10^{-8}$ for fiducial models (I) through (IV). 

\subsection{The aliasing effect} \label{subsec:aliasing}
In this section, we emphasize the importance of having {\it LISA} data in hand in the search for MBH binaries in LSST, by demonstrating the aliasing peaks that contaminate the periodogram when the light curve is not stretched using $\dot{f}$. 
Recall that we assume LSST's plan is to observe one pair of points 30 min apart every 6 nights (in a single filter). Since aliasing, in Fourier space, is a convolution of the (Fourier transforms of the) true signal and the time sampling, 
in the light curves where the time axes are not stretched (under the circumstance that $\dot{f}$ from GW is absent), false peaks are expected to occur near 
$\nu_{\rm Ny}$ (the so-called Nyquist frequency) and multiples of $\nu_{\rm Ny}$, where $\nu_{\rm Ny}=1/(2\pi \Delta t$) and $\Delta t$ is any combination of the spacings between data points. Aliasing affects models (I) and (II) the most because they have the highest orbital frequencies, a factor of few higher than the nominal (6 day)$^{-1}$ sampling frequency 
or the Nyquist frequency, $\nu_{\rm Ny}$ (half the sampling frequency). 

\begin{figure}
    \centering
    \includegraphics[width=\columnwidth]{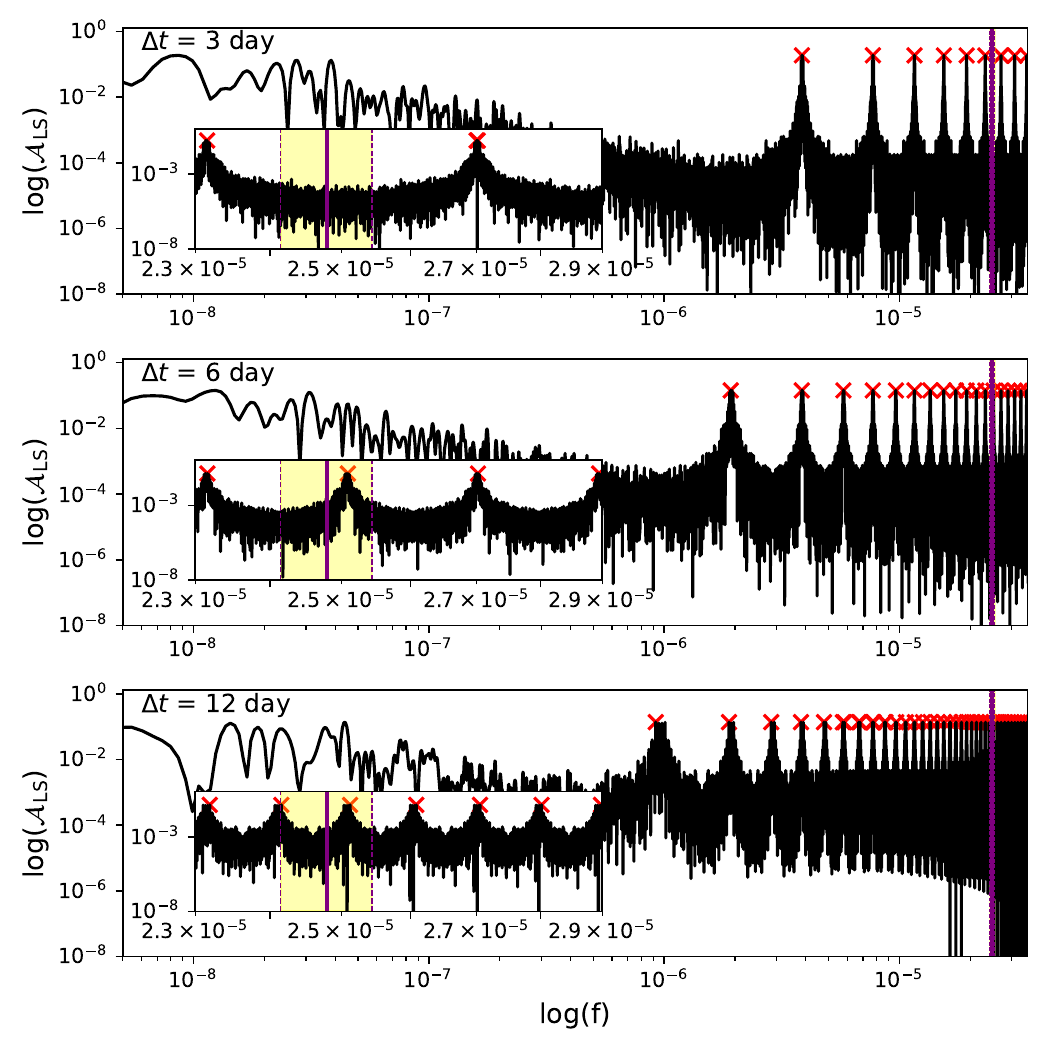}
    \caption{LS Periodograms of (unstretched) DRW light curves for a MBH binary with $M_{\rm bin}=3\times 10^5 {\rm M_{\odot}}$ and $z=0.3$, sampled at three different cadences, $\Delta{t}=3$d, 6d and 12d (top to bottom). We see local peaks (red crosses) that arise from pure noise and aliasing. The inserts show that, in this particular example, there is no false peak arising in the expected frequency range when the cadence is 3 days, but false peaks arise when the cadence is degraded to 6 days and 12 days. Generally, the LS periodogram becomes more noisy when the cadence becomes larger. }
    \label{fig:ls_cad}
\end{figure}

It is worth noting that at the time this paper is written, LSST aims to observe in six rotating filters, with slightly higher focus on some filters over the others. It is most likely that the cadence in the $i$-band will be approximately 6-12 days at best, with the possibility that some observations can be skipped due to weather conditions. However, it should be possible to combine data points from more than one band assuming color correlations between the bands, and to reduce the effective cadence to be shorter than 6 days. In Fig.~\ref{fig:ls_cad}, we compare the LS periodograms of the DRW light curves for binary model (I), using three different cadences -- the best cadence that LSST can achieve in a single filter, $\Delta t=3$ days (top panel), the fiducial cadence we adopt in this work, $\Delta t=6$ days (middle; this is the periodogram of the same light curve shown in Fig.~\ref{fig:drw_pg} if the $x$-coordinate is unstretched), and the ``worst" cadence, $\Delta t=12$ days (bottom panel), in the case that the $i$-band is not emphasized as much. Using the top panel as an example,
we see various peaks in the periodogram, located at frequencies corresponding to $\sim$ 1/(3d), 1/(30min), 1/(3d+30min), 1/(3d-30min), 1/(6d), and so on, as shown with red crosses. More importantly, there happens to be a false peak within the allowed frequency bin for $f({\rm -10~yr})$, when $\Delta t=6$ days and 12 days, as shown in the zoom-in insert.
From Fig.~\ref{fig:ls_cad}, it is apparent that many aliasing peaks arise when the time axis is not re-scaled using a known frequency evolution $f(x)$, and that the periodogram becomes more noisy as the cadence becomes larger. In reality, LSST sampling can be more uneven due to any skipped observations, leading to more random aliasing peaks in the periodogram. In principle, it is possible to remove aliasing peaks by identifying their locations using methods such as performing Monte Carlo simulations with real LSST cadences and obtaining distributions of $\mathit{A}_{\rm LS}$ at the aliasing frequencies. This is an approach we plan to analyze in future work. 

On the other hand, discovering a true peak in the periodogram of an LSST light curve, after the aliasing peaks are removed, can indicate the possibility that a genuine periodicity has been detected.  This then also means that the binary cannot be chirping significantly, since the peak would be washed out otherwise, see e.g. Fig~\ref{fig:mocklc_pg} where the (observed) orbital frequency changes by $\sim85\%$. The binary can have a slower chirp (smaller $\dot{f}$) if the total mass or the mass ratio is smaller. According to Eq.~\ref{eq:f_x}, if a genuine periodicity is detected for a slowly chirping binary, we can constrain the observed chirp mass of the binary (the total binary mass will be degenerate with the binary mass ratio and redshift). For example, let us assume that the periodogram can detect frequency chirps up to 10\%, i.e. $\delta{f}/f$ needs to be smaller than 0.1 over 10 years of LSST observation. This then implies that the redshifted binary chirp mass is $\log(\mathcal{M}_z/{\rm M_{\odot}})\leq 4.46$. At redshift $z=1 \pm 0.1$, this corresponds to binary chirp mass $\log(\mathcal{M}_c/{\rm M_{\odot}})\leq 4.16\pm 0.02$, or total mass, $\log(M_{\rm bin}/{\rm M_{\odot}}) \leq 4.81 \pm 0.02$ if the mass ratio is $q=0.1$. This mass is lower than the expected masses of ``verification binaries" above LSST's detection threshold.   We conclude that ``verification binaries" will most likely chirp more than 10\% during the 10-yr LSST observations, making it difficult to identify these binaries using only LSST data. 

\section{Results and Discussion} \label{sec:results}
Our main result is that for the four fiducial {\it LISA} binaries, there is no significant peak from pure noise that arises among the $100\times N_{\rm Q}$ realizations of DRW light curves, which means that their FAPs are less than $0.01/N_{\rm Q}$, corresponding to FAPs of $\lesssim 10^{-5}$, $\lesssim 10^{-6}$, $\lesssim 10^{-7}$ and $\lesssim 10^{-8}$, respectively. Therefore, all four binaries have an excellent chance to be uniquely identified in LSST, once their GWs are detected by {\it LISA}. 
A key ingredient in this analysis is that when the periodograms are constructed from the LSST light-curves, we take into account the expected chirp of the binaries, by re-scaling the time axes.  This re-scaling washes out false noise peaks while it allows the true periodicity to emerge. 
Identifying compact chirping binaries in the LSST data alone, without {\it LISA} information determining this re-scaling ab-initio, will be much more challenging. Due to the noise and aliasing effects, false peaks arise, which will likely cause higher FAP, even if some mitigation steps are taken to remove aliasing peaks. 
 
It is worthwhile to put these result in the context of other methods proposed to identify {\it LISA} MBH binaries or coalescencing MBHBs more generally. Namely, (i) several works have used cosmological simulations to identify the morphology and properties of the host galaxies associated with {\it LISA} binaries, and (ii) there have been several studies to search for periodicities in quasar light-curves in LSST (or in other large time domain surveys) that might be candidate MBHBs for GW detection in the {\it LISA} or PTA bands. 

The connection between coalescencing MBHBs in {\it LISA} or Pulsar Timing Arrays (PTAs) and the morphology of their host galaxies has been recently investigated in  cosmological simulations by \citet{Lops2023}, \citet{Izquierdo-Villalba2023} and \citet{Izquierdo-Villalba2023b}. They found through simulations that the number of host galaxies near a coalescencing {\it LISA} binary is too large (typically $>10^3$ galaxies) to allow unique identification.
Additionally, {\it LISA} MBHBs are typically found in low surface-brightness elliptical galaxies, making their morphological properties difficult to measure at redshifts $z>3$, even with LSST \citep{Izquierdo-Villalba2023b}. Even when observed and characterised at high significance, elliptical galaxies hosting single and binary MBHs of similar masses ($10^5-10^7 {\rm M_{\odot}}$) do not exhibit very different properties, such as the star-formation rate, gas fraction or metallicity. In conclusion, while complementary, it appears more profitable to search for {\it LISA} binaries using quasar observations rather than host galaxy observations. 
The latter approach, of course, is based on the assumption that coalescing binaries are bright quasars.
In practice, combining both types of identifications will very likely be helpful.

Recent works have also investigated different methods to search for periodic quasars associated with compact (super-)MBH binaries that are PTA precursors, using existing time-domain surveys, e.g. from the Catalina Real-Time Survey (CRTS) or mock LSST light curves \citep{Witt2022, Davis2023}. Others have studied the multi-messenger signatures of SMBHB candidates \citep{Xin2021a, Charisi2022}. 
These works focus on PTA presursors -- binaries with higher masses ($10^8-10^9{\rm M_{\odot}}$) and longer observed orbital periods \citep[typically one to a few years;][]{Graham+2015,Charisi+2016,Chen+2020}, compared to {\it LISA} binaries. Both \citet{Witt2022} and \citet{Davis2023} simulate mock LSST light curves in the i-band, but instead of focusing on the wide all-sky field (considered in this work), they use the observational criteria for the LSST deep drilling field.  
The former employs a Bayesian method and finds that the detectability of quasars are higher for shorter period, more variable (higher amplitude) and lower luminosity quasars, while their analysis is too computationally expensive to be done for a large sample of quasars. The latter arrives at similar conclusions, but they perform less expensive Monte Carlo simulations of binary light curves using a variety of binary parameters and DRW parameters. Note that the false positive probabilities computed in these works assume no pre-existing knowledge about the GW signal, which is the key novel ingredient in our analysis. We also emphasize that since {\it LISA} binaries have significantly shorter periods than PTA binaries, confirming a {\it LISA} binary will not require as long an observing baseline as confirming a PTA binary. However, one will face different challenges when identifying {\it LISA} binaries in LSST, without the confined frequency provided by {\it LISA}, including but not limited to the aliasing effects that we mention in \S~\ref{subsec:aliasing}.

An important outstanding question is the detectability, in practice, of the periodicity of ultra-compact MBHBs in the LSST all-sky survey. 
These binaries, which can be joint LSST and LISA sources, are much less massive and more compact than the ones investigated in \citet{Witt2022} and \citet{Davis2023}, making them more challenging to be identified due to the fact that they are fainter (making light-curves noisier and can necessitate stacking of exposures), their frequencies are higher (comparable or higher than the LSST cadence, complicating the analysis), and they also chirp by a significant fraction within the 10-year LSST baseline (which needs to be incorporated into the analysis as well).
We plan to pursue such analyses in our future work.

\section{Summary} \label{sec:summary}
In this work, we analyzed the prospects of identifying ultra-compact {\it LISA} MBH binaries as periodic quasar light curves in archival LSST data. We generated mock LSST light curves including a true binary component, exemplified by the sinusiodal variability expected from relativistic Doppler boost, as well as a stochastic quasar noise component described by the damped random walk (DRW) model and photometric errors. We quantified the false alarm probability at which the electromagnetic (EM) counterpart of the binary can be uniquely identified in LSST as the single quasar with significant periodicity matching the period expected based on the GW data. We performed this analysis for four fiducial {\it LISA} binaries with typical {\it LISA} source masses and redshifts, ($M_{\rm bin},z$)=($3\times10^5 {\rm M_{\odot}},0.3$), ($3\times10^6 {\rm M_{\odot}},0.3$), ($10^7 {\rm M_{\odot}},0.3$) and ($10^7 {\rm M_{\odot}}, 1$).  

Assuming that the GW signal associated with a MBH binary is detected by {\it LISA}, we extrapolate the quasar luminosity function to find the number $N_{\rm Q}$ of quasars in the LSST catalog with matching BH mass, redshift and sky location, with the errors implied by the {\it LISA} data. 
The BH masses and redshift are additionally uncertain due to (i) the uncertainty of the expected Eddington ratio
and LSST's photometric redshift.
The number of quasar counterpart candidates varies from $\sim 10^3$ to nearly $10^6$, driven primarily by the much poorer {\it LISA} sky location errors for more massive binaries (see Table~\ref{tab:model_summary}). 

Since the prospects of identifying a unique quasar counterpart based on mass and location are poor, we introduce the additional criterion that the quasar must display significant historical periodicity, at the frequencies expected from {\it LISA} data.  We compute these historical periods based on the expected chirp, along with their uncertainties in the {\it LISA} measurements. We then use Monte Carlo simulations of mock quasar light-curves to assess the false alarm probability (FAP) of finding significant periodicity by chance from pure noise.  Our analysis crucially incorporates the fact that the binary quasars chirp significantly already during the LSST observations (assumed to last 10 years).  Incorporating this chirp washed out "fake" noise-only peaks, and allows the true periodicity to emerge. Our main result is that this generally yields a single unique quasar with matching periodicity, with FAPs below 1 in $10^5$.

In our fiducial analysis, we adopted the {\it LISA} data characteristics (i.e. errors on the redshifted chirp mass, sky location, and distance) one week prior to merger. However, MBH binaries with lower masses and redshifts spend more time in the {\it LISA} band (e.g. see Fig. 4 in XH21) and generally have smaller uncertainties in localization, luminosity distance and chirp mass. Therefore, they can be detected as GWs and identified in LSST prior to one week before merger, and even up to 1 year before merger, although their ultra-short periodicity ($<1$d) can be harder to confirm because of aliasing. 
The scenario envisioned here therefore provides not just a novel way to uncover otherwise hard-to-identify EM counterparts to {\it LISA} sources, but also to facilitate 
follow-up observations during the last week of  (or even a longer period of time before) the merger, and look for additional evidence of binary signatures in tandem with the on-going GW signal.  These signatures include periodicity and chirping,  as well as EM signatures of de-coupling and merger 
in the UV and/or X-ray bands, to probe the very last stages of merger, further confirming the nature of the binary \citep{Krauth+2023, Franchini2024}.  

\section*{Acknowledgements}
We thank David Schiminovich for useful discussions, and David Izquierdo-Villalba and Alberto Mangiagli for providing clarifications about their results.
ZH gratefully acknowledges the hospitality of the Center for Computational Astrophysics (CCA) at the Flatiron Institute for a  sabbatical visit, where some of this work was performed.
ZH acknowledges support from NSF grant AST-2006176 and NASA grant 80NSSC22K0822.

\section*{Data availability}
No new data were generated or analysed in support of this research.

\bibliographystyle{mnras}
\bibliography{cx} 

\begin{thebibliography}{}
\makeatletter
\relax
\def\mn@urlcharsother{\let\do\@makeother \do\$\do\&\do\#\do\^\do\_\do\%\do\~}
\def\mn@doi{\begingroup\mn@urlcharsother \@ifnextchar [ {\mn@doi@}
  {\mn@doi@[]}}
\def\mn@doi@[#1]#2{\def\@tempa{#1}\ifx\@tempa\@empty \href
  {http://dx.doi.org/#2} {doi:#2}\else \href {http://dx.doi.org/#2} {#1}\fi
  \endgroup}
\def\mn@eprint#1#2{\mn@eprint@#1:#2::\@nil}
\def\mn@eprint@arXiv#1{\href {http://arxiv.org/abs/#1} {{\tt arXiv:#1}}}
\def\mn@eprint@dblp#1{\href {http://dblp.uni-trier.de/rec/bibtex/#1.xml}
  {dblp:#1}}
\def\mn@eprint@#1:#2:#3:#4\@nil{\def\@tempa {#1}\def\@tempb {#2}\def\@tempc
  {#3}\ifx \@tempc \@empty \let \@tempc \@tempb \let \@tempb \@tempa \fi \ifx
  \@tempb \@empty \def\@tempb {arXiv}\fi \@ifundefined
  {mn@eprint@\@tempb}{\@tempb:\@tempc}{\expandafter \expandafter \csname
  mn@eprint@\@tempb\endcsname \expandafter{\@tempc}}}

\bibitem[\protect\citeauthoryear{Andrae, Kim  \& Bailer-Jones}{Andrae
  et~al.}{2013}]{Andrae2013}
Andrae R.,  Kim D.~W.,   Bailer-Jones C.~A.,  2013, \mn@doi [\aap]
  {10.1051/0004-6361/201321335}, 554, 1

\bibitem[\protect\citeauthoryear{Ba{\~{n}}ados et~al.,}{Ba{\~{n}}ados
  et~al.}{2023}]{Banados2023}
Ba{\~{n}}ados E.,  et~al., 2023, \mn@doi [\apj, Supplement Series]
  {10.3847/1538-4365/acb3c7}, 265, 29

\bibitem[\protect\citeauthoryear{{Bortolas}, {Franchini}, {Bonetti}  \&
  {Sesana}}{{Bortolas} et~al.}{2021}]{Bortolas+2021}
{Bortolas} E.,  {Franchini} A.,  {Bonetti} M.,   {Sesana} A.,  2021, \mn@doi
  [\apjl] {10.3847/2041-8213/ac1c0c}, \href
  {https://ui.adsabs.harvard.edu/abs/2021ApJ...918L..15B} {918, L15}

\bibitem[\protect\citeauthoryear{{Burdge} et~al.,}{{Burdge}
  et~al.}{2020}]{Burdge+2020}
{Burdge} K.~B.,  et~al., 2020, \mn@doi [\apj] {10.3847/1538-4357/abc261}, \href
  {https://ui.adsabs.harvard.edu/abs/2020ApJ...905...32B} {905, 32}

\bibitem[\protect\citeauthoryear{Caplar, Lilly  \& Trakhtenbrot}{Caplar
  et~al.}{2017}]{Caplar2017}
Caplar N.,  Lilly S.~J.,   Trakhtenbrot B.,  2017, \mn@doi [\apj]
  {10.3847/1538-4357/834/2/111}, 834, 111

\bibitem[\protect\citeauthoryear{{Charisi}, {Bartos}, {Haiman}, {Price-Whelan},
  {Graham}, {Bellm}, {Laher}  \& {M{\'a}rka}}{{Charisi}
  et~al.}{2016}]{Charisi+2016}
{Charisi} M.,  {Bartos} I.,  {Haiman} Z.,  {Price-Whelan} A.~M.,  {Graham}
  M.~J.,  {Bellm} E.~C.,  {Laher} R.~R.,   {M{\'a}rka} S.,  2016, \mn@doi
  [\mnras] {10.1093/mnras/stw1838}, \href
  {http://adsabs.harvard.edu/abs/2016MNRAS.463.2145C} {463, 2145}

\bibitem[\protect\citeauthoryear{Charisi, Taylor, Runnoe, Bogdanovic  \&
  Trump}{Charisi et~al.}{2022}]{Charisi2022}
Charisi M.,  Taylor S.~R.,  Runnoe J.,  Bogdanovic T.,   Trump J.~R.,  2022,
  \mnras, 510, 5929

\bibitem[\protect\citeauthoryear{{Chen} et~al.,}{{Chen}
  et~al.}{2020}]{Chen+2020}
{Chen} Y.-C.,  et~al., 2020, \mn@doi [\mnras] {10.1093/mnras/staa2957}, \href
  {https://ui.adsabs.harvard.edu/abs/2020MNRAS.499.2245C} {499, 2245}

\bibitem[\protect\citeauthoryear{Cocchiararo, Franchini, Lupi  \&
  Sesana}{Cocchiararo et~al.}{2024}]{Cocchiararo2024}
Cocchiararo F.,  Franchini A.,  Lupi A.,   Sesana A.,  2024, \apj, submitted;
  e-print arXiv:2402.05175

\bibitem[\protect\citeauthoryear{{Colpi} et~al.,}{{Colpi}
  et~al.}{2024}]{LISA2024}
{Colpi} M.,  et~al., 2024, ESA LISA Definition Study Report; e-print
  arXiv:2402.07571, \href
  {https://ui.adsabs.harvard.edu/abs/2024arXiv240207571C} {}

\bibitem[\protect\citeauthoryear{Corrales, Haiman  \& MacFadyen}{Corrales
  et~al.}{2010}]{Corrales2010}
Corrales L.~R.,  Haiman Z.,   MacFadyen A.,  2010, \mn@doi [\mnras]
  {10.1111/j.1365-2966.2010.16324.x}, 404, 947

\bibitem[\protect\citeauthoryear{{D'Orazio}, {Haiman}  \&
  {Schiminovich}}{{D'Orazio} et~al.}{2015}]{Dorazio+2015}
{D'Orazio} D.~J.,  {Haiman} Z.,   {Schiminovich} D.,  2015, \mn@doi [\nat]
  {10.1038/nature15262}, \href
  {http://adsabs.harvard.edu/abs/2015Natur.525..351D} {525, 351}

\bibitem[\protect\citeauthoryear{Davis et~al.,}{Davis et~al.}{2023}]{Davis2023}
Davis M.~C.,  et~al., 2023, \apj, submitted; e-print arXiv:2311.10851v1

\bibitem[\protect\citeauthoryear{Dittmann, Ryan  \& Miller}{Dittmann
  et~al.}{2023}]{Dittmann2023}
Dittmann A.~J.,  Ryan G.,   Miller M.~C.,  2023, \mn@doi [\apjl]
  {10.3847/2041-8213/acd183}, 949, L30

\bibitem[\protect\citeauthoryear{Duffell, D'Orazio, Derdzinski, Haiman,
  MacFadyen, Rosen  \& Zrake}{Duffell et~al.}{2020}]{Duffell2020}
Duffell P.~C.,  D'Orazio D.,  Derdzinski A.,  Haiman Z.,  MacFadyen A.,  Rosen
  A.~L.,   Zrake J.,  2020, \mn@doi [\apj] {10.3847/1538-4357/abab95}, 901, 25

\bibitem[\protect\citeauthoryear{Fan et~al.,}{Fan et~al.}{2006}]{Fan2006}
Fan X.,  et~al., 2006, \mn@doi [\apj] {10.1086/504836}, 132, 117

\bibitem[\protect\citeauthoryear{{Farris}, {Duffell}, {MacFadyen}  \&
  {Haiman}}{{Farris} et~al.}{2014}]{Farris+2014}
{Farris} B.~D.,  {Duffell} P.,  {MacFadyen} A.~I.,   {Haiman} Z.,  2014,
  \mn@doi [\apj] {10.1088/0004-637X/783/2/134}, \href
  {http://adsabs.harvard.edu/abs/2014ApJ...783..134F} {783, 134}

\bibitem[\protect\citeauthoryear{{Farris}, {Duffell}, {MacFadyen}  \&
  {Haiman}}{{Farris} et~al.}{2015}]{Farris+2015}
{Farris} B.~D.,  {Duffell} P.,  {MacFadyen} A.~I.,   {Haiman} Z.,  2015,
  \mn@doi [\mnras] {10.1093/mnrasl/slu184}, \href
  {http://adsabs.harvard.edu/abs/2015MNRAS.447L..80F} {447, L80}

\bibitem[\protect\citeauthoryear{Franchini, Prato, Longarini  \&
  Sesana}{Franchini et~al.}{2024}]{Franchini2024}
Franchini A.,  Prato A.,  Longarini C.,   Sesana A.,  2024, Astronomy \&
  Astrophysics, submitted; e-print arXiv:2402.00938

\bibitem[\protect\citeauthoryear{{Graham} et~al.,}{{Graham}
  et~al.}{2015}]{Graham+2015}
{Graham} M.~J.,  et~al., 2015, \mn@doi [\mnras] {10.1093/mnras/stv1726}, \href
  {http://adsabs.harvard.edu/abs/2015MNRAS.453.1562G} {453, 1562}

\bibitem[\protect\citeauthoryear{Haiman, Kocsis  \& Menou}{Haiman
  et~al.}{2009}]{Haiman2009a}
Haiman Z.,  Kocsis B.,   Menou K.,  2009, \mn@doi [\apj]
  {10.1088/0004-637X/700/2/1952}, 700, 1952

\bibitem[\protect\citeauthoryear{Hopkins \& Hernquist}{Hopkins \&
  Hernquist}{2009}]{Hopkins2009}
Hopkins P.~F.,  Hernquist L.,  2009, \mn@doi [\apj]
  {10.1088/0004-637X/698/2/1550}, 698, 1550

\bibitem[\protect\citeauthoryear{Ivezic \& MacLeod}{Ivezic \&
  MacLeod}{2013}]{Ivezic2013}
Ivezic Z.,  MacLeod C.~L.,  2013, \mn@doi [Proceedings of the International
  Astronomical Union] {10.1017/S1743921314004396}, 9, 395

\bibitem[\protect\citeauthoryear{Izquierdo-Villalba, Sesana  \&
  Colpi}{Izquierdo-Villalba et~al.}{2023a}]{Izquierdo-Villalba2023}
Izquierdo-Villalba D.,  Sesana A.,   Colpi M.,  2023a, \mn@doi [\mnras]
  {10.1093/mnras/stac3677}, 519, 2083

\bibitem[\protect\citeauthoryear{Izquierdo-Villalba, Colpi, Volonteri, Spinoso,
  Bonoli  \& Sesana}{Izquierdo-Villalba
  et~al.}{2023b}]{Izquierdo-Villalba2023b}
Izquierdo-Villalba D.,  Colpi M.,  Volonteri M.,  Spinoso D.,  Bonoli S.,
  Sesana A.,  2023b, \mn@doi [\aap] {10.1051/0004-6361/202347008}, 677, 1

\bibitem[\protect\citeauthoryear{Katz, Kelley, Dosopoulou, Berry, Blecha  \&
  Larson}{Katz et~al.}{2020}]{Katz2020}
Katz M.~L.,  Kelley L.~Z.,  Dosopoulou F.,  Berry S.,  Blecha L.,   Larson
  S.~L.,  2020, \mn@doi [\mnras] {10.1093/mnras/stz3102}, 491, 2301

\bibitem[\protect\citeauthoryear{Kelly, Bechtold  \& Siemiginowska}{Kelly
  et~al.}{2009}]{Kelly2009}
Kelly B.~C.,  Bechtold J.,   Siemiginowska A.,  2009, \mn@doi [\apj]
  {10.1088/0004-637X/698/1/895}, 698, 895

\bibitem[\protect\citeauthoryear{Kochanek et~al.,}{Kochanek
  et~al.}{2012}]{Kochanek2012}
Kochanek C.~S.,  et~al., 2012, \mn@doi [\apj, Supplement Series]
  {10.1088/0067-0049/200/1/8}, 200

\bibitem[\protect\citeauthoryear{{Kollmeier} et~al.,}{{Kollmeier}
  et~al.}{2006}]{Kollmeier+2006}
{Kollmeier} J.~A.,  et~al., 2006, \mn@doi [\apj] {10.1086/505646}, \href
  {https://ui.adsabs.harvard.edu/abs/2006ApJ...648..128K} {648, 128}

\bibitem[\protect\citeauthoryear{{Koz{\l}owski} et~al.,}{{Koz{\l}owski}
  et~al.}{2010}]{Kozlowski2010}
{Koz{\l}owski} S.,  et~al., 2010, \mn@doi [\apj] {10.1088/0004-637X/708/2/927},
  \href {https://ui.adsabs.harvard.edu/abs/2010ApJ...708..927K} {708, 927}

\bibitem[\protect\citeauthoryear{{Krauth}, {Davelaar}, {Haiman},
  {Westernacher-schneider}, {Zrake}  \& {Macfadyen}}{{Krauth}
  et~al.}{2023}]{Krauth+2023}
{Krauth} L.~M.,  {Davelaar} J.,  {Haiman} Z.,  {Westernacher-schneider} J.~R.,
  {Zrake} J.,   {Macfadyen} A.,  2023, \mn@doi [\mnras]
  {10.1093/mnras/stad3095}, 526, 5441

\bibitem[\protect\citeauthoryear{Kulkarni, Worseck  \& Hennawi}{Kulkarni
  et~al.}{2019}]{Kulkarni2019}
Kulkarni G.,  Worseck G.,   Hennawi J.~F.,  2019, \mn@doi [\mnras]
  {10.1093/mnras/stz1493}, 31, 1035

\bibitem[\protect\citeauthoryear{{LSST Science Collaboration} et~al.,}{{LSST
  Science Collaboration} et~al.}{2009}]{LSSTScienceCollaboration2009}
{LSST Science Collaboration} et~al., 2009, arXiv e-prints

\bibitem[\protect\citeauthoryear{Lops, Izquierdo-Villalba, Colpi, Bonoli,
  Sesana  \& Mangiagli}{Lops et~al.}{2023}]{Lops2023}
Lops G.,  Izquierdo-Villalba D.,  Colpi M.,  Bonoli S.,  Sesana A.,   Mangiagli
  A.,  2023, \mn@doi [\mnras] {10.1093/mnras/stad058}, 519, 5962

\bibitem[\protect\citeauthoryear{MacLeod et~al.,}{MacLeod
  et~al.}{2010}]{MacLeod2010}
MacLeod C.~L.,  et~al., 2010, \mn@doi [\apj] {10.1088/0004-637X/721/2/1014},
  721, 1014

\bibitem[\protect\citeauthoryear{Mangiagli et~al.,}{Mangiagli
  et~al.}{2020}]{Mangiagli2020}
Mangiagli A.,  et~al., 2020, \mn@doi [\prd] {10.1103/PhysRevD.102.084056}, 102,
  84056

\bibitem[\protect\citeauthoryear{Milosavljevi{\'{c}} \&
  Phinney}{Milosavljevi{\'{c}} \& Phinney}{2005}]{Milosavljevic2005}
Milosavljevi{\'{c}} M.,  Phinney E.~S.,  2005, \mn@doi [\apj] {10.1086/429618},
  622, L93

\bibitem[\protect\citeauthoryear{{Peters} \& {Mathews}}{{Peters} \&
  {Mathews}}{1963}]{Peters1963}
{Peters} P.~C.,  {Mathews} J.,  1963, Physical Review, 131, 435

\bibitem[\protect\citeauthoryear{Rossi, Lodato, Armitage, Pringle  \&
  King}{Rossi et~al.}{2010}]{Rossi2010}
Rossi E.~M.,  Lodato G.,  Armitage P.~J.,  Pringle J.~E.,   King A.~R.,  2010,
  \mn@doi [\mnras] {10.1111/j.1365-2966.2009.15802.x}, 401, 2021

\bibitem[\protect\citeauthoryear{{Tang}, {Haiman}  \& {MacFadyen}}{{Tang}
  et~al.}{2018}]{Tang+2018}
{Tang} Y.,  {Haiman} Z.,   {MacFadyen} A.,  2018, \mn@doi [\mnras]
  {10.1093/mnras/sty423}, \href
  {http://adsabs.harvard.edu/abs/2018MNRAS.476.2249T} {476, 2249}

\bibitem[\protect\citeauthoryear{{Vaughan}, {Uttley}, {Markowitz},
  {Huppenkothen}, {Middleton}, {Alston}, {Scargle}  \& {Farr}}{{Vaughan}
  et~al.}{2016}]{Vaughan2016}
{Vaughan} S.,  {Uttley} P.,  {Markowitz} A.~G.,  {Huppenkothen} D.,
  {Middleton} M.~J.,  {Alston} W.~N.,  {Scargle} J.~D.,   {Farr} W.~M.,  2016,
  \mn@doi [\mnras] {10.1093/mnras/stw1412}, \href
  {http://adsabs.harvard.edu/abs/2016MNRAS.461.3145V} {461, 3145}

\bibitem[\protect\citeauthoryear{{Vera C. Rubin Observatory Project Science
  Team Committee}}{{Vera C. Rubin Observatory Project Science Team
  Committee}}{2023}]{SurveyCadence}
{Vera C. Rubin Observatory Project Science Team Committee} 2023, pp 1--60

\bibitem[\protect\citeauthoryear{Westernacher-Schneider, Zrake, MacFadyen  \&
  Haiman}{Westernacher-Schneider et~al.}{2022}]{Westernacher-Schneider2022}
Westernacher-Schneider J.~R.,  Zrake J.,  MacFadyen A.,   Haiman Z.,  2022,
  \mn@doi [\prd] {10.1103/physrevd.106.103010}, 106

\bibitem[\protect\citeauthoryear{Willott et~al.,}{Willott
  et~al.}{2010}]{Willott2010}
Willott C.~J.,  et~al., 2010, \mn@doi [\aj] {10.1088/0004-6256/139/3/906}, 139,
  906

\bibitem[\protect\citeauthoryear{Witt, Charisi, Taylor  \& Burke-Spolaor}{Witt
  et~al.}{2022}]{Witt2022}
Witt C.~A.,  Charisi M.,  Taylor S.~R.,   Burke-Spolaor S.,  2022, \mn@doi
  [\apj] {10.3847/1538-4357/ac8356}, 936, 89

\bibitem[\protect\citeauthoryear{Xin \& Haiman}{Xin \& Haiman}{2021}]{Xin2021}
Xin C.,  Haiman Z.,  2021, \mn@doi [\mnras] {10.1093/mnras/stab1856}, 506, 2408

\bibitem[\protect\citeauthoryear{Xin, Mingarelli  \& Hazboun}{Xin
  et~al.}{2021}]{Xin2021a}
Xin C.,  Mingarelli C. M.~F.,   Hazboun J.~S.,  2021, \mn@doi [\apj]
  {10.3847/1538-4357/ac01c5}, 915, 97

\bibitem[\protect\citeauthoryear{Yang et~al.,}{Yang et~al.}{2019}]{Yang2019}
Yang J.,  et~al., 2019, \mn@doi [\aj] {10.3847/1538-3881/ab1be1}, 157, 236

\bibitem[\protect\citeauthoryear{{Zrake}, {Tiede}, {MacFadyen}  \&
  {Haiman}}{{Zrake} et~al.}{2021}]{Zrake+2021}
{Zrake} J.,  {Tiede} C.,  {MacFadyen} A.,   {Haiman} Z.,  2021, \mn@doi [\apjl]
  {10.3847/2041-8213/abdd1c}, \href
  {https://ui.adsabs.harvard.edu/abs/2021ApJ...909L..13Z} {909, L13}

\makeatother
\end{thebibliography}
\bsp
\label{lastpage}
\end{document}